\documentclass[review,12pt,sort&compress]{elsarticle}

\usepackage{amssymb}

\usepackage{color} 
\usepackage{float} 
\usepackage{subfig}
\usepackage{booktabs}
\usepackage{tabularx} 
\usepackage{subfiles} 
\usepackage{multirow}
\usepackage{hyperref}
\usepackage{enumitem}
\usepackage{diagbox}
\usepackage{makecell}
\usepackage{epstopdf}

\journal{Mechanics Research Communications}

\begin{document}

\begin{frontmatter}



\title{Miniaturized gas-solid fluidized beds
\tnoteref{label_note_copyright} \tnoteref{label_note_doi}
}

\tnotetext[label_note_copyright]{\copyright 2023. This manuscript version is made available under the CC-BY-NC-ND 4.0 license http://creativecommons.org/licenses/by-nc-nd/4.0/}

\tnotetext[label_note_doi]{Accepted Manuscript for Mechanics Research Communications, v. 131, 104146, 2023, DOI:10.1016/j.mechrescom.2023.104146}


\author[label_david]{Fernando D. C\'u\~nez}
\author[label_erick]{Erick M. Franklin\corref{cor1}}
\ead{erick.franklin@unicamp.br}
\cortext[cor1]{Corresponding author. phone: +55 19 35213375}

\affiliation[label_david]{organization={Department of Earth and Environmental Sciences, University of Rochester},
	addressline={227 Hutchison Hall,}, 
	city={Rochester},
	postcode={14611}, 
	state={NY},
	country={USA}}

\affiliation[label_erick]{organization={School of Mechanical Engineering, University of Campinas - UNICAMP},
            addressline={Rua Mendeleyev, 200}, 
            city={Campinas},
            postcode={13083-860}, 
            state={SP},
            country={Brazil}}


\begin{abstract}
Fluidized beds are suspensions of grains by ascending fluids in tubes, and are commonly used in industry given their high rates of mass and heat transfers between the solids and fluid. Although usually employed in large scales (tube diameter of the order of the meter), fluidized beds have a great potential in much smaller scales (order of the millimeter) for processes involving powder and fluids. Of particular interest is the pharmaceutical industry, which can take advantage of mm-scale fluidized beds for promoting diffusion of species, classifying grains, or peeling individual particles. This paper reports experiments with a mm-scale gas-solid fluidized bed, which consisted of 0.5-mm-diameter glass particles suspended by an air flow in a 3-mm-ID glass tube. We filmed the bed with a high-speed camera and processed the images with a numerical code for tracking both the entire bed and individual particles. We found that instabilities in the form of alternating high- and low-compactness regions (known respectively as plugs and bubbles) appear in the bed, and that the fluctuating energy of particles (known as granular temperature) is relatively low within plugs. Therefore, mm-scale beds have much reduced agitation and transfer rates when compared to their m-scale counterparts. We show also that increasing the flow velocity does not avoid the appearance of plugs, though the granular temperature increases, mitigating the problem. Our results shed light on detailed mechanisms taking place within the miniaturized bed, providing insights for chemical and pharmaceutical processes involving powders.
\end{abstract}




\end{frontmatter}


\section{Introduction}

Fluidized beds consist in a suspension of grains by an ascending fluid in a tube, being commonly used in industry because of their high rates of mass and heat transfers between the solids and fluid. They are usually employed in large scale facilities (tube diameter of the order of meters) for processes involving combustion, coating, drying, and catalytic and non-catalytic reactions, for instance, in industries as diverse as mining, oil, food and energy, among others. However, fluidized beds have a great potential, still poorly explored, in much smaller scales (order of millimeters) for processes involving powders and fluids and needing relatively fast mass and/or heat transfers \cite{Zhang, Han2}. Previous works on mm-scale and cm-scale fluidized beds, called micro fluidized beds (MFBs \cite{Zhang, Qie}), showed that MFBs can be successfully employed for mechanical and chemical processes. For example, they have been used for particle encapsulation \cite{Schreiber, Rodriguez}, pyrolysis \cite{Jia, Gao, Mao, Yu}, catalytic cracking \cite{Boffito, Guo2}, gasification \cite{Zeng, Zhang2, Cortazar}, capture of $CO_2$ \cite{Fang, Shen}, bioproduction \cite{Wu, Liu4, Pereiro}, and wastewater treatment \cite{Kuyukina, Kwak}. Despite the advantages of using MFBs for process intensification, commercial applications are still in their inception \cite{Zhang}. To date, some examples of mature products are the gas-solid MFB analytical device presented by Han et al. \cite{Han}, the  MFB thermogravimetric analyzer developed by Samih et al. \cite{Samih}, and the compact MFBs proposed by Li et al. \cite{Li3} for $CO_2$ capture.

One case of particular interest is the pharmaceutical industry, which deals with powders ranging from 1 to 500 $\mu$m that must be classified, segregated, mixed, compressed, peeled, or undergo surface diffusion (with a given fluid) in order to produce tablets and pills \cite{Kornblum, Katstra, Fung, Ervasti, Azad, Azad2, Shi}. In addition, vaccines have recently been searched in the powder form \cite{Jiang, Huang, Gomez, Heida}, including those for diseases highly impacting human activities such as influenza and COVID-19 \cite{Amorij, Heida}. Vaccines in powder form are considered more stable and of easier logistics (dispensing cold-chain facilities) \cite{Amorij}, allowing for faster distribution in cases of pandemic outbreaks. Therefore, the pharmaceutical industry can take advantage of mm-scale fluidized beds for promoting faster diffusion of species, particle classification, or particle peeling, to cite just a few processes, for powder processing.

Although of relatively simple construction, fluidized beds have a rich dynamics, with different particle-fluid, particle-particle, and particle-wall interactions happening at the same time, and being susceptible to distinct patterns and flow regimes \cite{Nicolas, Sundaresan, guazzelli2004fluidized}. The different regimes for usual (large) beds were extensively investigated over the last decades \cite{Miller, Geldart, Rietema, Menon, Escudero}, but few works investigated smaller beds with a ratio between the tube $D$ and grain $d$ diameters within 10 and 100 \cite{Anderson, ElKaissy, Didwania, Zenit, Zenit2, Duru, Duru2, Goldman, Aguilar, Ghatage}. In especial, Goldman and Swinney \cite{Goldman} showed that beds with $D/d$ $\lessapprox$ 100 submitted to fluid decelerations and slight accelerations are subject to crystallization and jamming. Crystallization corresponds to a static lattice of high compactness and with small fluctuations of grains (absence only of macroscopic motion), occurring at fluid velocities slightly above that for minimum fluidization, $U_{mf}$, whereas jamming corresponds to a static lattice where even small fluctuations of individual particles are not present (absence of microscopic motion). Usually, jamming occurs after increasing slightly the fluid velocity in an already crystallized bed. One of their main findings was that bed crystallization has similarities with equilibrium glass transition, one of them being the dependence on the deceleration rate. 

There are still fewer investigations on very-narrow beds, those for which $D/d$ $\lesssim$ 10, all of them concerning solid-liquid fluidized beds (SLFBs). In C\'u\~nez and Franklin \cite{Cunez}, we identified the appearance of instabilities in the form of alternating high- and low-compactness regions, called plugs and bubbles, and showed that they are caused by dense networks of contact forces that percolate within the bed and reach the tube wall. This is a characteristic of high confinement, and appears only in the very-narrow case. Later, we \cite{Cunez2} investigated the segregation and the inversion of granular layers in the case of beds consisting of bidisperse mixtures. We found the trajectories of individual particles and the characteristic times for inversion in the very-narrow case, showing that the network of contact forces and confinement greatly affect layer inversion.

More recently, we \cite{Cunez3} studied the behavior of very-narrow SLFBs under partial de-fluidization and re-fluidization, similar to Goldman and Swinney \cite{Goldman}, but varying grain types (and thus $U_{mf}$). We found that crystallization of portions of the bed can happen with fluid velocities above $U_{mf}$, as in Ref. \cite{Goldman}, but that different lattice structures, however, appear depending on the grain type. Still, we showed a relative independence of crystallization on the deceleration rate, and that, by increasing the flow velocity, jamming intensity depends on the particle type. Afterward, we \cite{Cunez4} investigated the problem further by inquiring into plug formation, crystallization and jamming in very-narrow beds consisting of bonded spheres (duos and trios). We showed that there are different structures within the bed and distinct motions for duos and trios, and that jamming can occur suddenly for trios, calling into question the fluidization conditions for these cases.

There has been a recent interest in MFBs, but investigations of mm-scale beds remain scarce. Guo et al. \cite{Guo} investigated fluidization conditions in gas-solid MFBs of varying sizes (tube inner diameter varying from 4.3 to 25.5 mm), and found that the pressure drop is relatively small and the minimum fluidization velocity high when compared to larger beds and classical correlations, which they explained based on an increase in bed voidage. Do Nascimento et al. \cite{Nascimento} investigated liquid-solid MFBs (hydraulic diameters of 1 and 2 mm), and showed that adhesion forces can reach values of the same order of hydrodynamic and gravitational forces, providing a new explanation for the increase in the minimum fluidization velocity. The results were corroborated later by Li et al. \cite{Li2}, who showed that the expansion of solid-liquid MFBs have low expansion ratios and higher minimum velocities than large-scale beds. Other studies on MFBs \cite{Rao, Wang3, Doroodchi} showed the same tendency. However, in spite of the increasing interest in MFBs over the years, previous studies were concerned mainly with macroscopic aspects such as the pressure drop, coefficients of heat transfer and minimum fluidization velocities. To the authors' knowledge, there is a lack of information concerning the microscopic scale, e.g., the formation of bed structures and the level of fluctuations at the grain scale. These data are fundamental to determine regimes of higher mass and heat transfers, elaborate adequate correlations, and validate numerical simulations.

Despite the recent progresses on very-narrow beds and MFBs, their behaviors are far from being completely understood. For instance, previous studies on MLBs were exclusively on the bed scale, and those on very-narrow beds dealt exclusively with solid-liquid beds, so that the behavior of less dense fluids, where Archimedes, lubrication and added mass forces can be negligible, has never been investigated. In this paper we investigate experimentally a gas-solid mm-scale MFB in the very-narrow case, the bed consisting of 0.5-mm-diameter glass particles submitted to different ascending air flows in a 3-mm-ID glass tube. We filmed the bed with a high-speed camera and processed the images with a numerical code written in the course of this work for tracking both the entire bed (macroscopic scale) and individual particles (microscopic scale). We found that instabilities in the form of alternating high- and low-compactness regions (plugs and bubbles) appear in the bed, and that the fluctuating energy of particles (known as granular temperature) is relatively low within plugs. Therefore, mm-scale beds have much reduced agitation and transfer rates when compared to their m-scale counterparts. We also show that increasing the flow velocity does not avoid the appearance of plugs, though the granular temperature increases, mitigating the problem. Our results shed light on the mechanics taking place within the miniaturized bed, and is of interest for chemical and pharmaceutical processes involving powders and fluids and needing relatively fast mass and/or heat transfers.

In the following, Sec. \ref{sec:setup} describes the experimental setup, Sec. \ref{sec:results} presents the results and their discussion, and Sec. \ref{sec:conclusions} concludes the paper.


\section{Experimental setup}
\label{sec:setup}

\begin{figure}[ht]
	\centering
	\includegraphics[width=0.75\columnwidth]{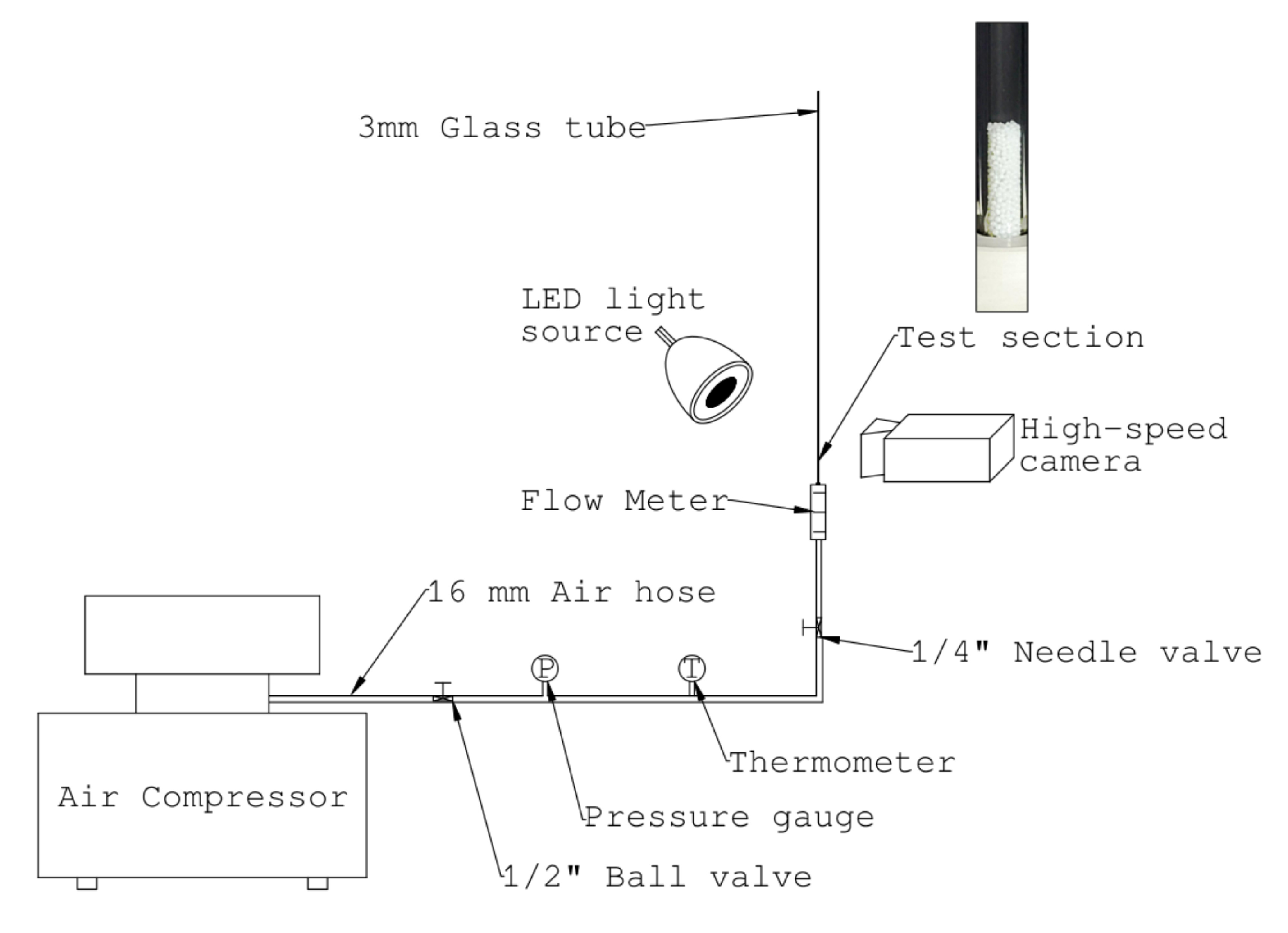}
	\caption{Layout of the experimental setup (test section shown in the inset).}
	\label{fig_layout}
\end{figure}

The experimental setup consisted basically of an air compressor mounted on a pressure vessel, controlling vanes, a pressure transducer, a termocouple,  a flow meter, a flow homogenizer, and a 3-mm-ID vertical tube. The vertical tube was 1 m long and made of glass (borosilicate), and it was aligned vertically within $\pm 2^{\circ}$. Downstream the tube, air discharged at atmosphere pressure. Figure \ref{fig_layout} shows a layout of the experimental setup and a photograph of part of the glass tube (in the inset). A photograph of the experimental apparatus is available in the Supplementary Material.

Controlled grains were settled in the test section, forming a granular bed, and upward air flows were imposed by adjusting the controlling vanes. We used 0.5-mm-diameter glass spheres with density $\rho_p$ = 2500 kg/m$^3$ for the beds (a microscopy image of the used spheres is available in the Supplementary Material). We investigated beds with initial heights $h_{if}$ of 7, 8.6 and 10.4 mm, under cross-sectional mean velocities of air (superficial velocities) $\overline{U}$ equal to 1.4, 1.8, 2.0 and 2.2 m/s. Therefore, $D/d$ = 6, and the numbers of Stokes and Reynolds based on terminal velocities are $St_t \,=\, v_t d \rho_p / (9\mu_f)$  = 2.86 $\times$ 10$^4$ and $Re_t \,=\, \rho_f  v_t d / \mu_f$ = 1.26 $\times$ 10$^2$, respectively, where $v_t$ is the terminal velocity of one single particle and $\mu_f$ is the dynamic viscosity of the fluid. The inception of fluidization was determined based on acquired images by detecting the initial motion of grains and considering the respective air velocity as the incipient fluidization velocity $U_{if}$. As pointed out in Zhang et al. \cite{Zhang}, the velocity for minimum fluidization can be determined from image analysis by identifying the inception of bed expansion. In the case of MFBs, the value thus obtained can differ from the minimum fluidization velocity obtained from pressure drop, $U_{mf}$, since the high friction and adhesion of MFBs engender considerable hysteresis between fluidization and de-fluidization. In addition, MFBs present large deviations from correlations for regular beds \cite{Zhang}, so that they are not used here. For the inception, we found $U_{if}$ = 1.2 m/s, and we consider the particle fraction $\phi_0$ $\approx$ 0.5 based on C\'u\~nez and Franklin \cite{Cunez3}. The settling velocity was estimated based on the Richardson--Zaki correlation, $v_s = v_t \left( 1-\phi_0 \right) ^{2.4}$, and found equal to 0.70 m/s. The parameters varied in the tests are summarized in Tab. \ref{tab_exp}, which presents also the Reynolds numbers for both tube and grains based on the superficial velocity, $Re_D \,=\, \rho_f  \overline{U} d / \mu_f$ and $Re_d \,=\, \rho_f  \overline{U} d / \mu_f$, respectively. Values of $St_t$ and $Re_t$ indicate that grains have considerable inertia with respect to the employed fluid, and values of $Re_D$ that the base flow (without solid particles) is laminar.

\begin{table}[h!]
	\begin{center}
		\caption{Summary of tested conditions: case label, initial height $h_{if}$, superficial velocity $\overline{U}$, and Reynolds numbers for the tube and grains based on the superficial velocity, $Re_D$ and $Re_d$, respectively.}
		\begin{tabular}{c c c c c}
			\hline\hline
			Case  & $h_{if}$ & $\overline{U}$ & $Re_D$ & $Re_d$ \\
			$\cdots$  & mm & m/s & $\cdots$ & $\cdots$\\
			\hline
			a & 7 & 1.4 & 283 & 47\\
			b & 7 & 1.8 & 363 & 61\\
			c & 7 & 2.0 & 404 & 67\\
			d & 7 & 2.2 & 444 & 74\\
			e & 8.6 & 1.4 & 283 & 47\\
			f & 8.6 & 1.8 & 363 & 61\\
			g & 8.6 & 2.0 & 404 & 67\\
			h & 8.6 & 2.2 & 444 & 74\\
			i & 10.4 & 1.4 & 283 & 47\\
			j & 10.4 & 1.8 & 363 & 61\\
			k & 10.4 & 2.0 & 404 & 67\\
			l & 10.4 & 2.2 & 444 & 74\\
			\hline
		\end{tabular}
		\label{tab_exp}
	\end{center}
\end{table}

\begin{sloppypar}
A high-speed camera of complementary metal-oxide-semiconductor (CMOS) type with maximum resolution of 2560 px $\times$ 1600 px at 800 Hz was placed perpendicularly to the test section and acquired images of the beds. The region of interest (ROI) was set to 2560 px $\times$ 120 px at 2000 Hz, and the field of view was of 132 mm $\times$ 6.2 mm, corresponding to approximately 19.4 px/mm. A lens of $60$ mm focal distance and F2.8 maximum aperture was mounted on the camera, and we made use of lamps of light-emitting diode (LED) branched to a continuous-current source to avoid beating between the camera frequency and light source.
\end{sloppypar}


\section{Results and discussion}
\label{sec:results}

As soon as the air flow started, we observed the bed expansion with the formation of alternating high- and low-compactness regions (granular plugs and gas bubbles, respectively) occupying the entire tube cross section. These forms, which are similar to those obtained for very-narrow SLFBs at larger scales (cm scale) \cite{Cunez, Cunez2, Cunez3, Cunez4}, were nearly one dimensional and propagated upward with characteristic lengths and celerities. However, different from cm-scale SLFBs, we observed neither crystallization nor jamming. In order to visualize the bed structures, we placed  side by side snapshots from experiments and show two examples in Fig. \ref{fig_snap_exp}. Movies from experiments, and snapshots of the remaining cases are available in the Supplementary Material and in an open repository \cite{Supplemental3}.

\begin{figure}
	\begin{center}
			\includegraphics[width=0.95\columnwidth]{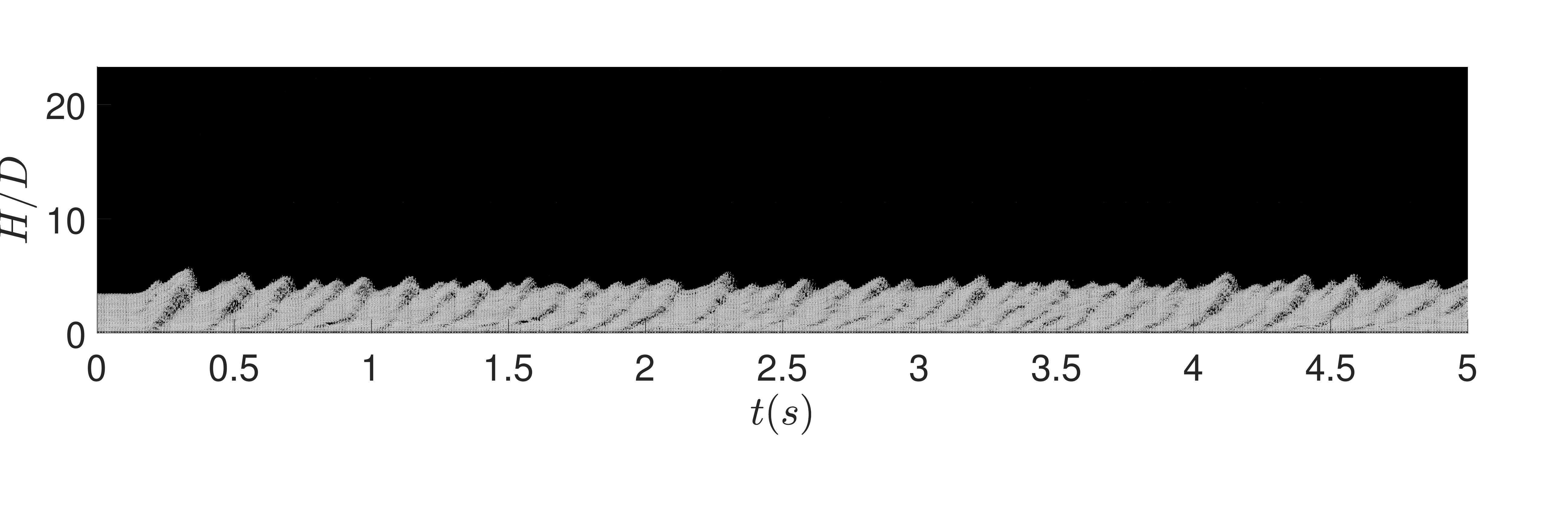}\\
			(a)\\
			\includegraphics[width=0.95\columnwidth]{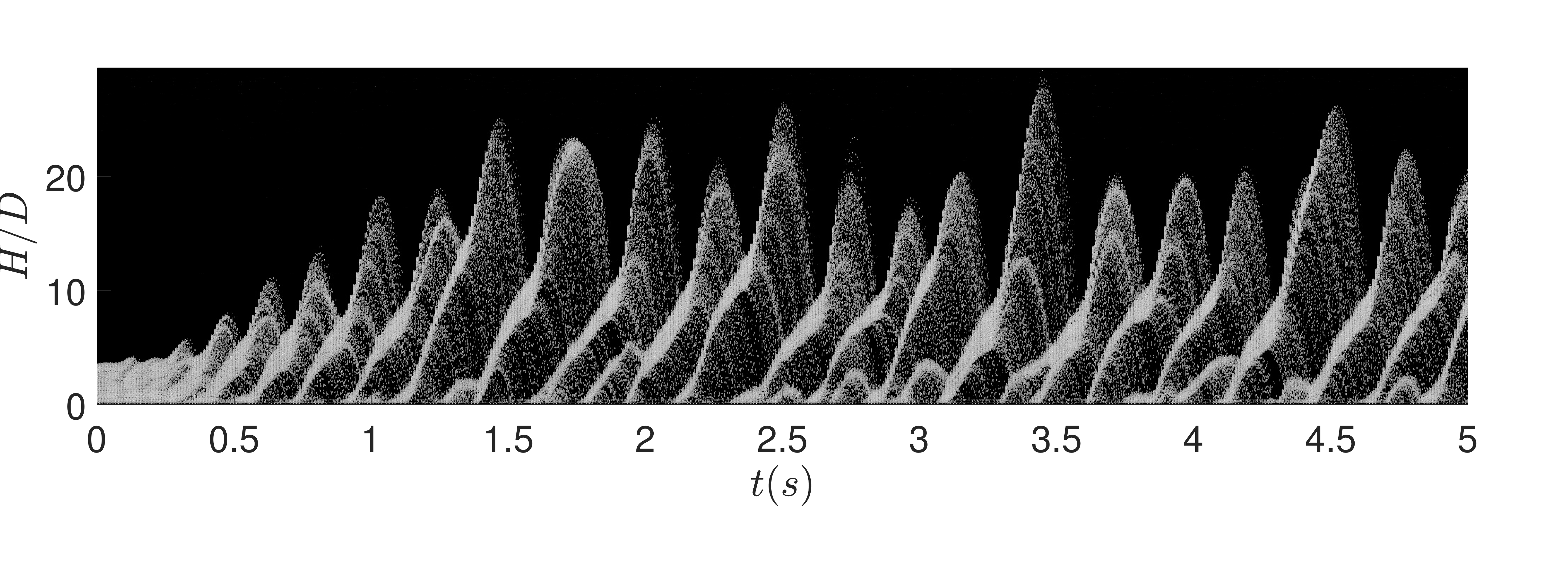}\\
			(b)\\
	\end{center}
	\caption{Snapshots from experiments placed side by side for cases \textit{i} and \textit{k} of Tab. \ref{tab_exp}. The abscissa shows the time $t$ and the ordinate the bed height normalized by the tube diameter, $H/D$. Time between frames is of 10$^{-2}$ s. Movies from the experiments corresponding to these snapshots are available in the Supplementary Material.}
	\label{fig_snap_exp}
\end{figure}

We detected and tracked the different structures (plugs and bubbles) and the top of the bed in the high-speed movies by using numerical scripts written in the course of this work and based on Refs. \cite{Cunez, Cunez2, Cunez3, Cunez4} (the numerical scripts used for image processing are available in an open repository \cite{Supplemental3}). With that, we computed the characteristic lengths and celerities of the bed at macroscopic level, and present their average values in Tab. \ref{tab_macro} (dimensional forms and the corresponding standard deviations are available in the Supplementary Material). Table \ref{tab_macro} presents the dimensionless values of the mean height of beds $H_{av}/h_{if}$, average upward and downward celerities of the top of beds, $c_{up}/\overline{U}$ and $c_{down}/\overline{U}$, respectively, frequency of oscillation of the bed top  $f\,d/\overline{U}$, average lengths of plugs, $\lambda / D$, and average celerity of plugs $c_{plug}/\overline{U}$ ($f$ is normalized by the characteristic time for the settling of particles, $d/\overline{U}$). $\lambda$ and $c_{plug}$ were obtained by computing time and ensemble averages of the length and celerity of plugs, respectively.

\begin{table}[h!]
	\begin{center}
		\caption{Macroscopic parameters: case, mean height of beds $H_{av}$ normalized by $h_{if}$, average upward and downward celerities of the bed top, $c_{up}$ and $c_{down}$, respectively, normalized by $\overline{U}$, frequency of oscillation of the bed top $f$ normalized by the settling time of particles $d/\overline{U}$, average lengths of plugs normalized by the tube diameter, $\lambda / D$, and average celerity of plugs $c_{plug}$ normalized by $\overline{U}$.}
		\begin{tabular}{c c c c c c c}
			\hline\hline
			Case & $H_{av}/h_{if}$ & $c_{up}/\overline{U}$ & $c_{down}/\overline{U}$ & $f\,d/\overline{U}$ & $\lambda / D$ & $c_{plug}/\overline{U}$\\
			\hline
			a & 1.34 & 0.018 & -0.020 & 2.6 & 3.05 & 0.012\\
			b & 2.26 & 0.046 & -0.057 & 1.7 & 2.39 & 0.023\\
			c & 4.61 & 0.095 & -0.114 & 1.0 & 1.14 & 0.039\\
			d & 6.14 & 0.099 & -0.112 & 1.0 & 0.97 & 0.049\\
			e & 1.38 & 0.025 & -0.033 & 2.4 & 3.63 & 0.015\\
			f & 3.64 & 0.084 & -0.098 & 1.3 & 1.24 & 0.042\\
			g & 4.67 & 0.092 & -0.104 & 1.1 & 0.94 & 0.044\\
			h & 8.57 & 0.138 & -0.145 & 0.7 & 0.72 & 0.065\\
			i & 1.23 & 0.020 & -0.027 & 2.1 & 4.11 & 0.011\\
			j & 2.84 & 0.082 & -0.096 & 1.5 & 1.41 & 0.036\\
			k & 4.34 & 0.104 & -0.123 & 1.1 & 1.02 & 0.048\\
			l & 8.32 & 0.135 & -0.140 & 0.9 & 0.67 & 0.063\\
			\hline
		\end{tabular}
		\label{tab_macro}
	\end{center}
\end{table}

\begin{figure}[h!]
	\begin{center}
		\begin{minipage}{0.45\linewidth}
			\begin{tabular}{c}
				\includegraphics[width=0.85\linewidth]{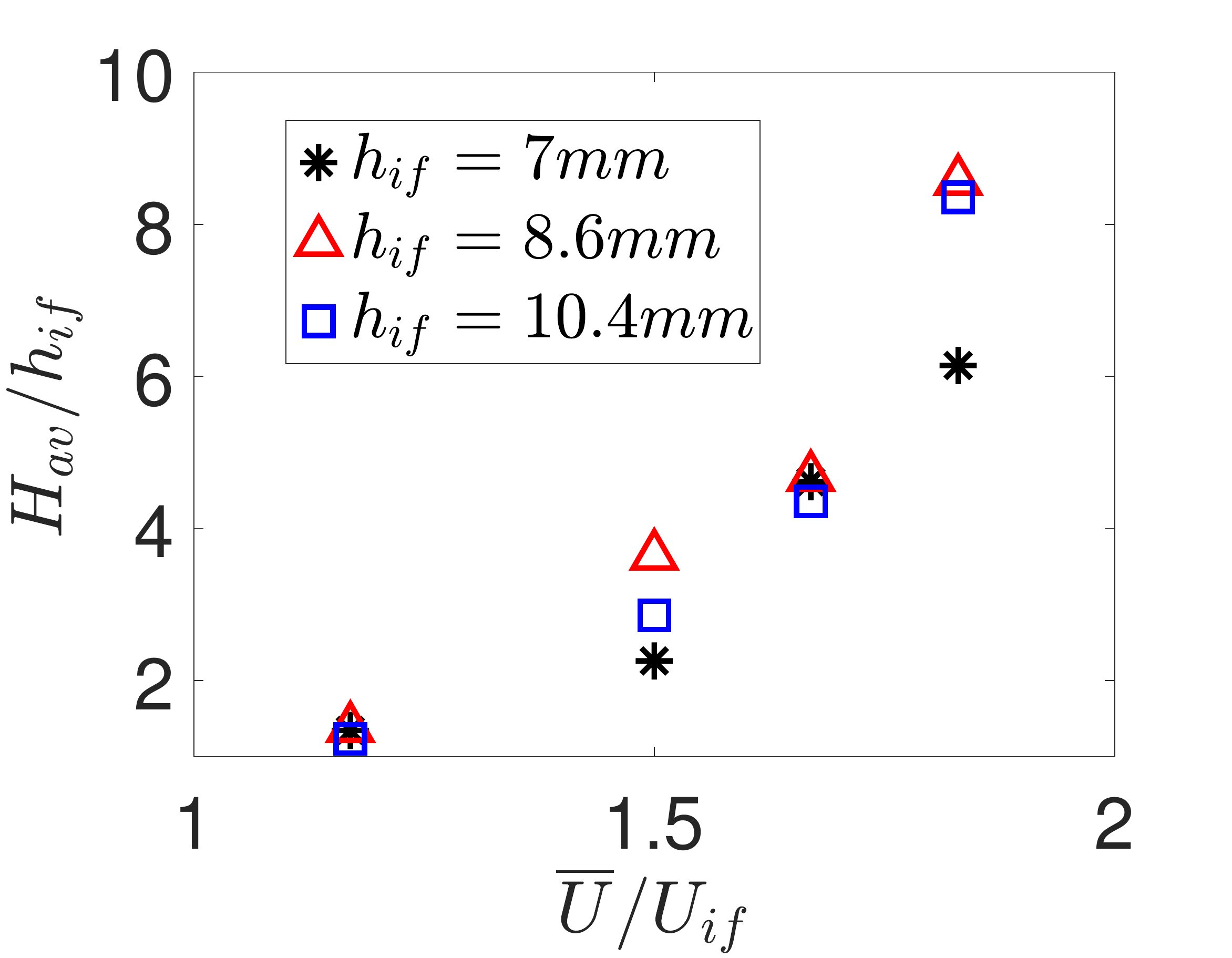}\\
				(a)
			\end{tabular}
		\end{minipage}
		\hfill
		\begin{minipage}{0.45\linewidth}
			\begin{tabular}{c}
				\includegraphics[width=0.85\linewidth]{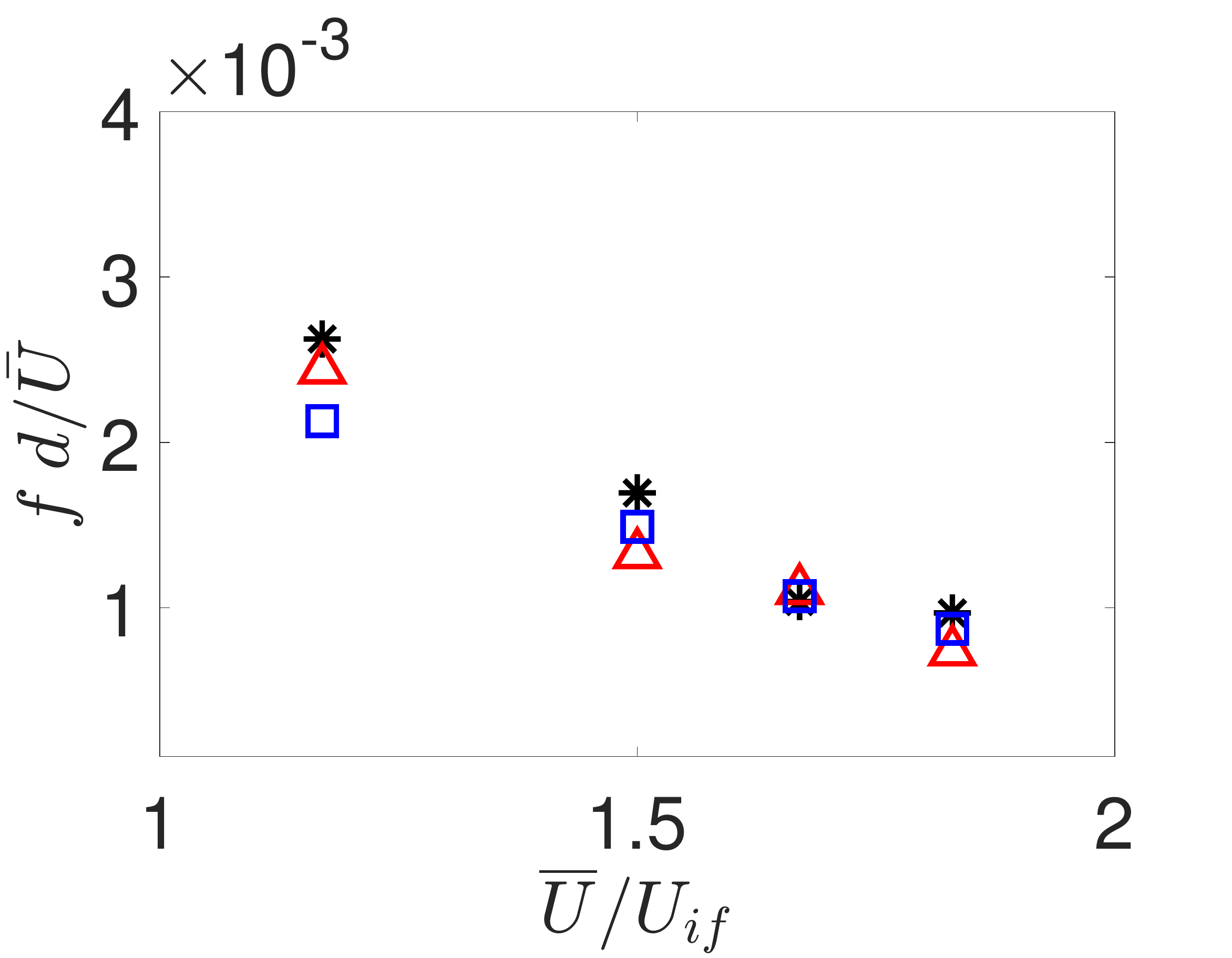}\\
				(b)
			\end{tabular}
		\end{minipage}
		\hfill
		\begin{minipage}{0.45\linewidth}
			\begin{tabular}{c}
				\includegraphics[width=0.85\linewidth]{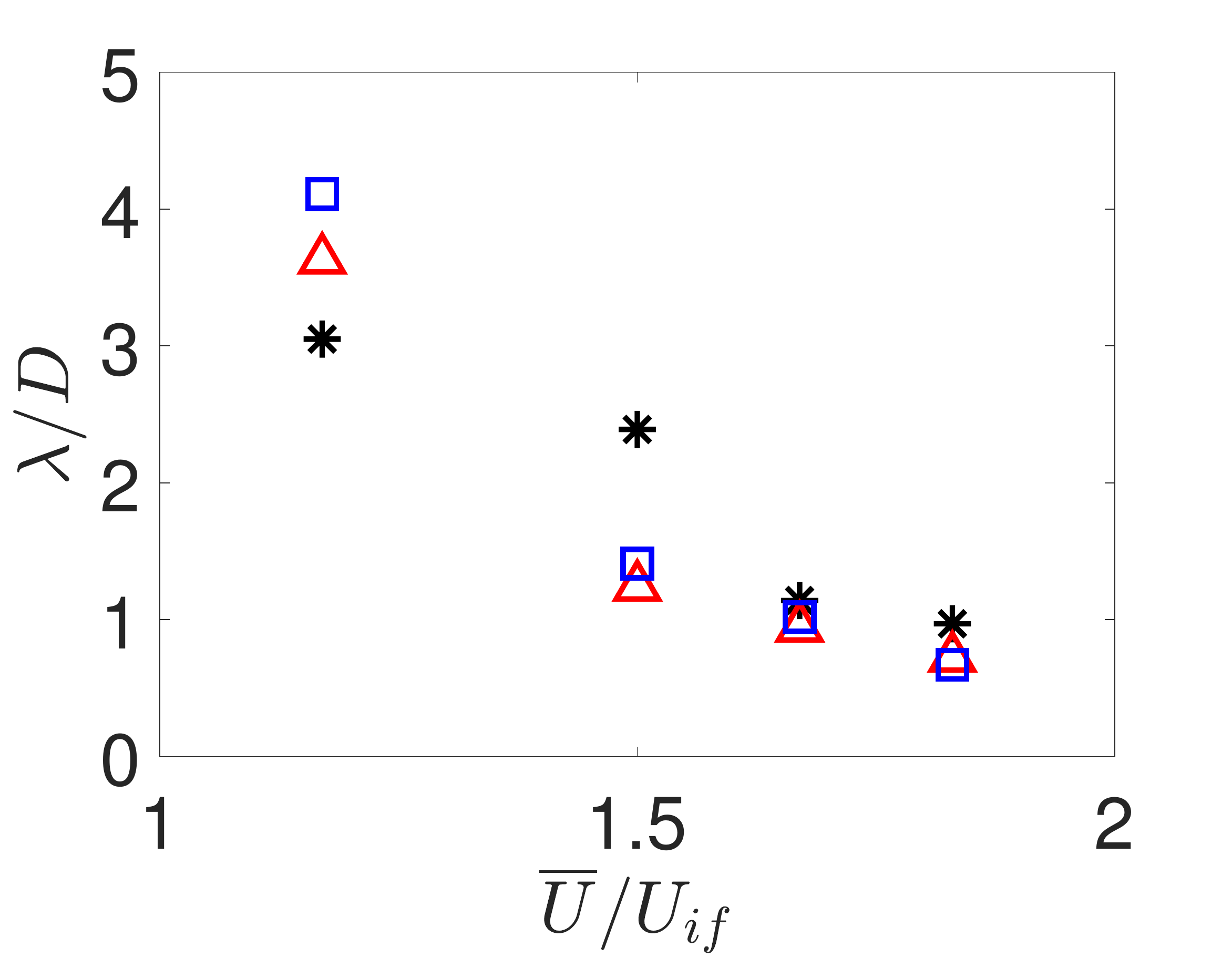}\\
				(c)
			\end{tabular}
		\end{minipage}
		\begin{minipage}{0.45\linewidth}
			\begin{tabular}{c}
				\includegraphics[width=0.85\linewidth]{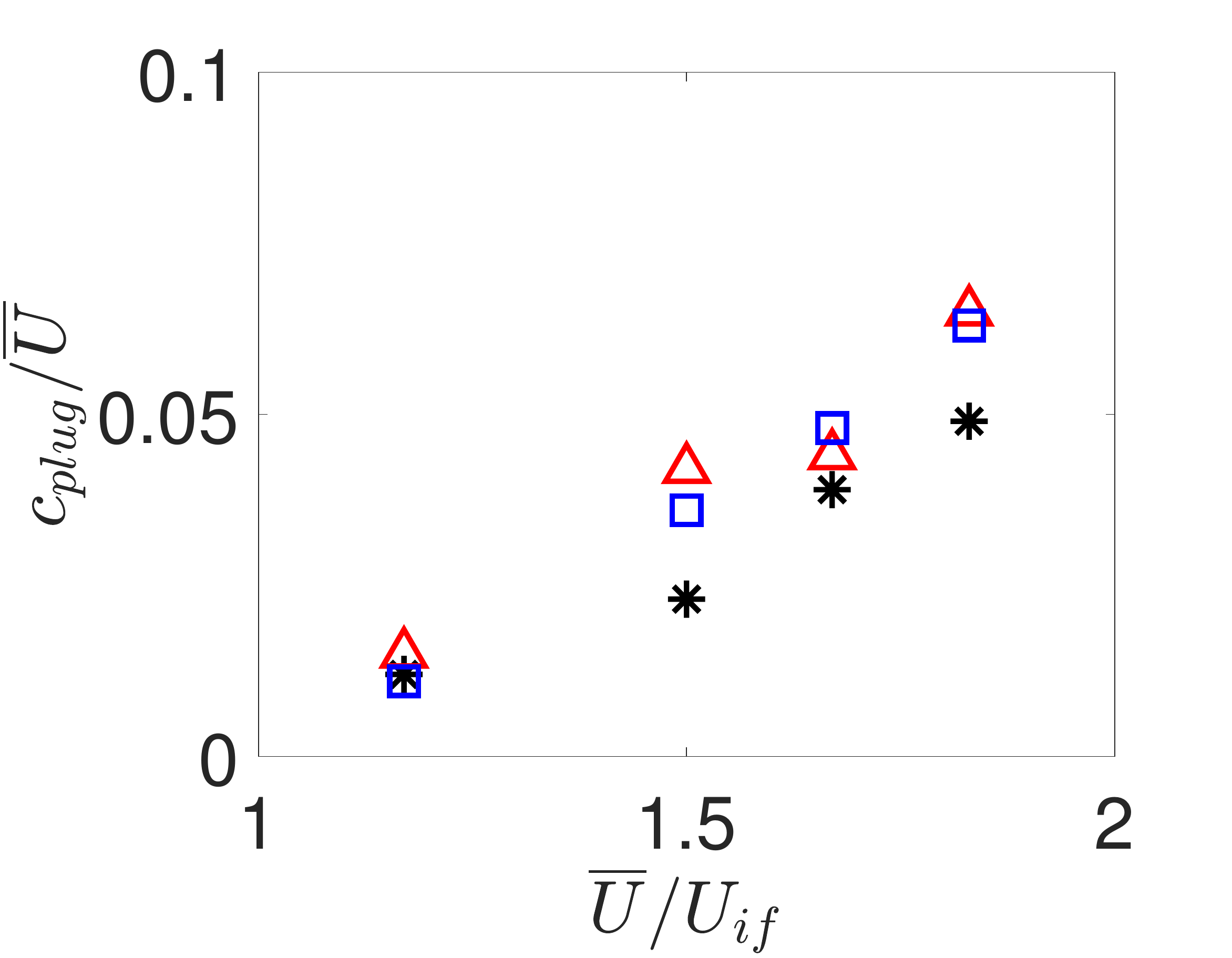}\\
				(d)
			\end{tabular}
		\end{minipage}
		\hfill
	\end{center}
	\caption{(a) Average height of the bed $H_{av}/h_{if}$, (b) frequency of oscillation of the bed top $fd/\overline{U}$, (c) plug length $\lambda / D$ and plug celerity $c_{plug}/\overline{U}$ as functions of the mean velocity $\overline{U} / U_{if}$ of air, parameterized by $h_{if}$.}
	\label{fig:height_freq}
\end{figure}

With the imposed air flow, plugs and bubbles propagated upward and made the top of the bed oscillate between minimum and maximum extremes for a given air velocity. We computed the mean bed height $H_{av}$ as the average between minimum and maximum values of the bed height $H$, and the upward and downward celerities of the top, $c_{up}$ and $c_{down}$, as the derivative of the instantaneous height during bed expansion and contraction, respectively. As observed from the snapshots, by increasing the fluid velocity the bed presents higher values of $H$, reaching thus higher expansion values. For the beds investigated, augmenting $\overline{U}$ from 1.4 to 2.2 m/s (an increase of 57\%) increased $H_{av}$ by 4-7 times. This can be seen in Fig. \ref{fig:height_freq}a, which shows $H/D$ as a function of $\overline{U}/U_{if}$.  With the increase in the bed expansion, the upward and downward celerities reach higher magnitudes and show strong variations with the air velocity: for the same augmentation in $\overline{U}$, both $c_{up}$ and $c_{down}$ increase by one order of magnitude (7 to 11 times in the dimensional values and 4 to 7 times in the dimensionless values).

For the plug length $\lambda$, it decreases with increasing the air flow, with a small dependence on the initial height of the bed, as shown in Fig. \ref{fig:height_freq}c. A similar trend was observed in the case of very-narrow SLFBs \cite{Cunez, Cunez2, Cunez3, Cunez4}. In the present case, $\lambda$ decreased by 70--85\% when  $\overline{U}$ is increased by 57\%. Concerning the plug celerities, they increase as $\overline{U}$ increases, with the average celerity $c_{plug}$ increasing 6--9 times ($c_{plug}/\overline{U}$ increasing 4--6 times) for the same 57\% increase in $\overline{U}$. This is shown in Fig. \ref{fig:height_freq}d, from which we also observe that the dependence on the bed height is small.

In addition to the just described, we observe two general behaviors of the bed. For smaller values of $\overline{U}$, as in Fig. \ref{fig_snap_exp}a, the plugs are large and oscillate with one main frequency and relatively few plug mergings. For higher values of $\overline{U}$, as in Fig. \ref{fig_snap_exp}b, plugs are smaller and we can observe two main frequencies and a considerable number of plug mergings (from plug-plug collisions). The second frequency appears basically from plug mergings, occurring mainly at the bottom of the bed (such as in $t$ $\approx$ 2 s at $H/D$ $\approx$ 5), which results in smaller frequencies of oscillation of the bed top, but keeping the average celerity of plugs $c_{plug}$ high. From Tab. \ref{tab_macro} and Fig. \ref{fig:height_freq}b, we observe that the frequency of oscillation decreases with the fluid velocity and is roughly independent of the bed height, while Fig. \ref{fig:height_freq}c shows that the plug length varies with the bed height for small fluid velocities.

\begin{figure}[ht]
	\centering
	\includegraphics[width=0.75\columnwidth]{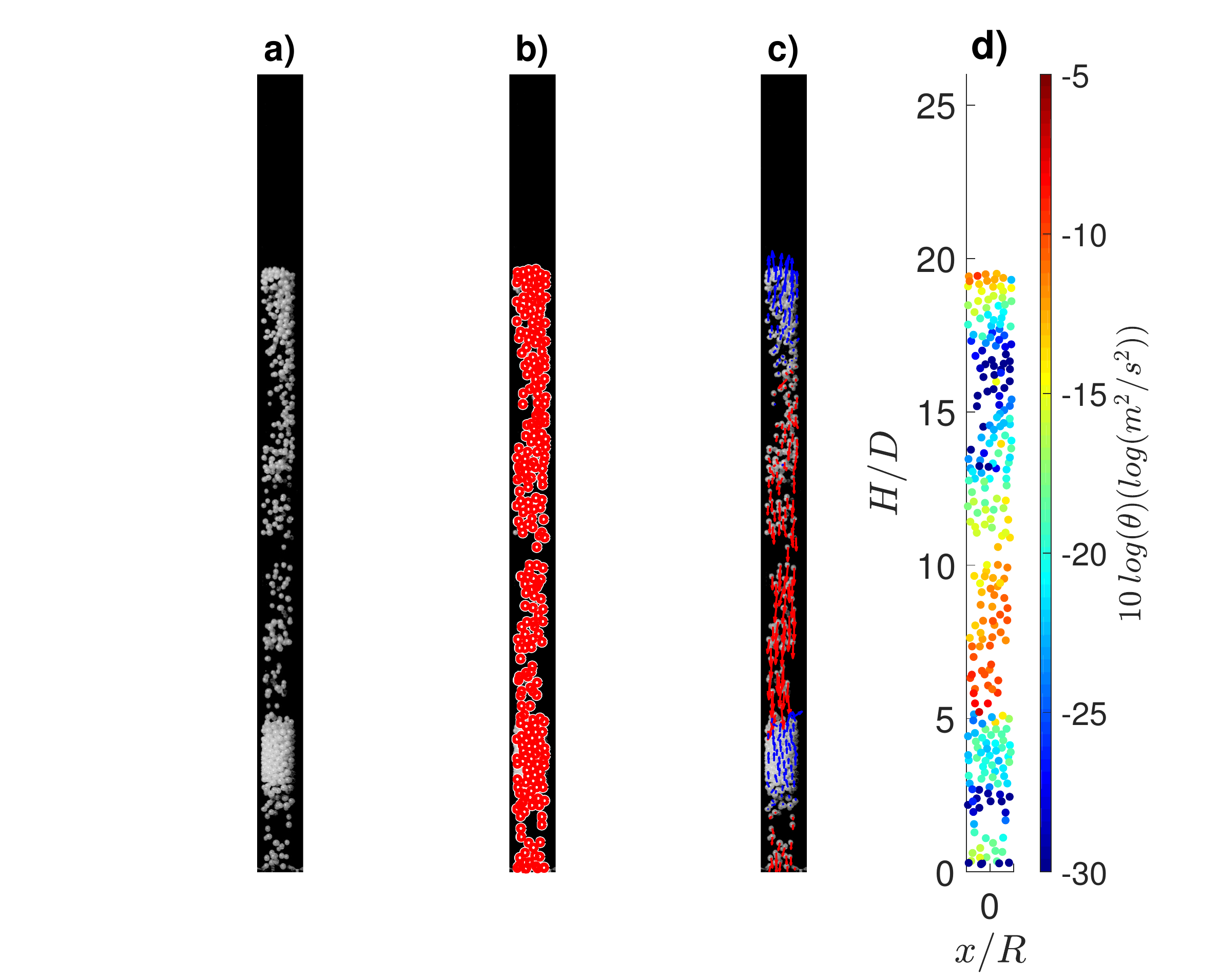}
	\caption{Example of image processing: (a) raw image; (b) particle identification; (c) velocity of each particle; (d) granular temperature of each particle.}
	\label{fig_image_processing}
\end{figure}

Because neither crystallization, jamming nor electrostatic effects were noticed, and based on the bed expansion, plug lengths, celerities, and plug-plug collisions observed, the use of higher air velocities seems to imply stronger mass and heat transfers in the MFBs investigated, something of great interest for industrial processes. In order to investigate that further, we measured the fluctuations at the grain scale. For that, we tracked along images the grains that were visible, which correspond to those in contact with the tube wall, by using numerical scripts written in the course of this work and based on Kelley and Ouellette \cite{Kelley} and Houssais et al. \cite{Houssais_1}. Figure \ref{fig_image_processing} shows the sequence followed by the image processing that we used: raw image (Fig. \ref{fig_image_processing}a), identification of particles (Fig. \ref{fig_image_processing}b), velocity of each particle (Fig. \ref{fig_image_processing}c), and granular temperature of each particle (explained next, Fig. \ref{fig_image_processing}d). Although images correspond to a frontal view of a cylindrical plane, we considered the motion of tracked grains as occurring in a Cartesian coordinate system. Other than missing some of the grains that are not in contact with the frontal wall, the use of a Cartesian plane implies in missing motions perpendicular to such plane and adding parallax errors (the latter mitigated by the use of a visualization box).

\begin{figure}
	\begin{center}
		\begin{tabular}{c}
			\includegraphics[width=0.7\columnwidth]{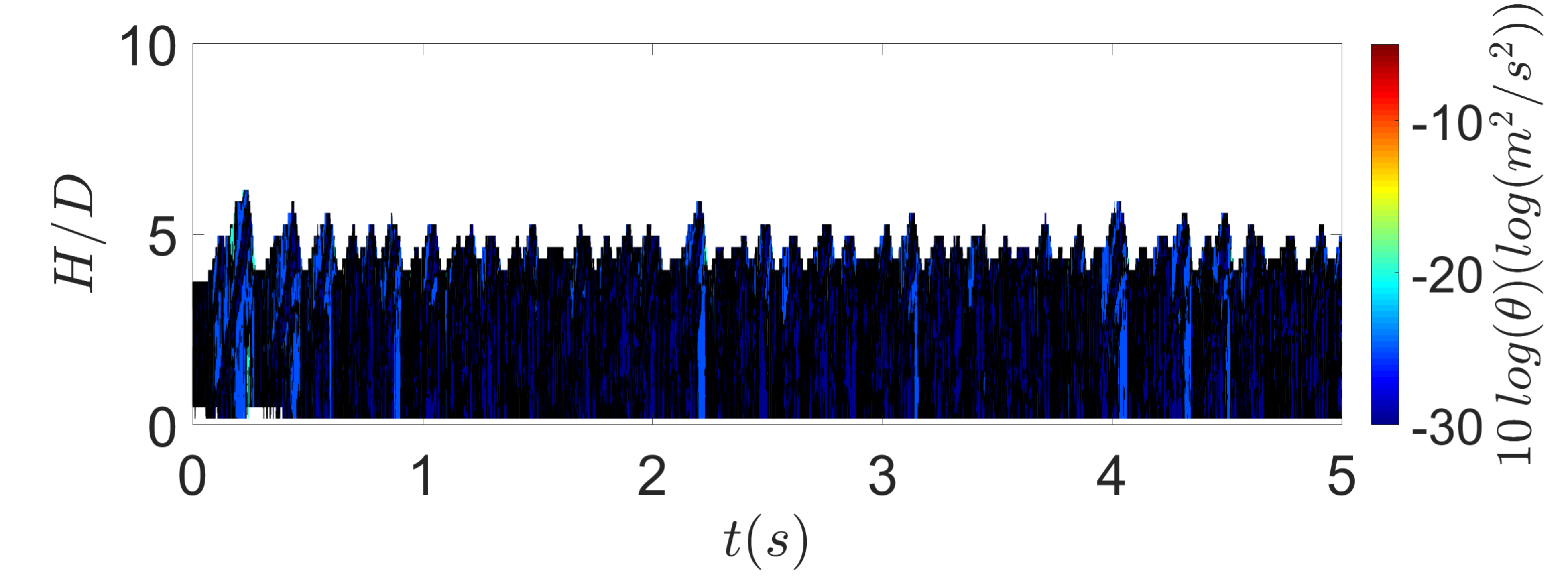}\\
			(a)\\
			\includegraphics[width=0.7\columnwidth]{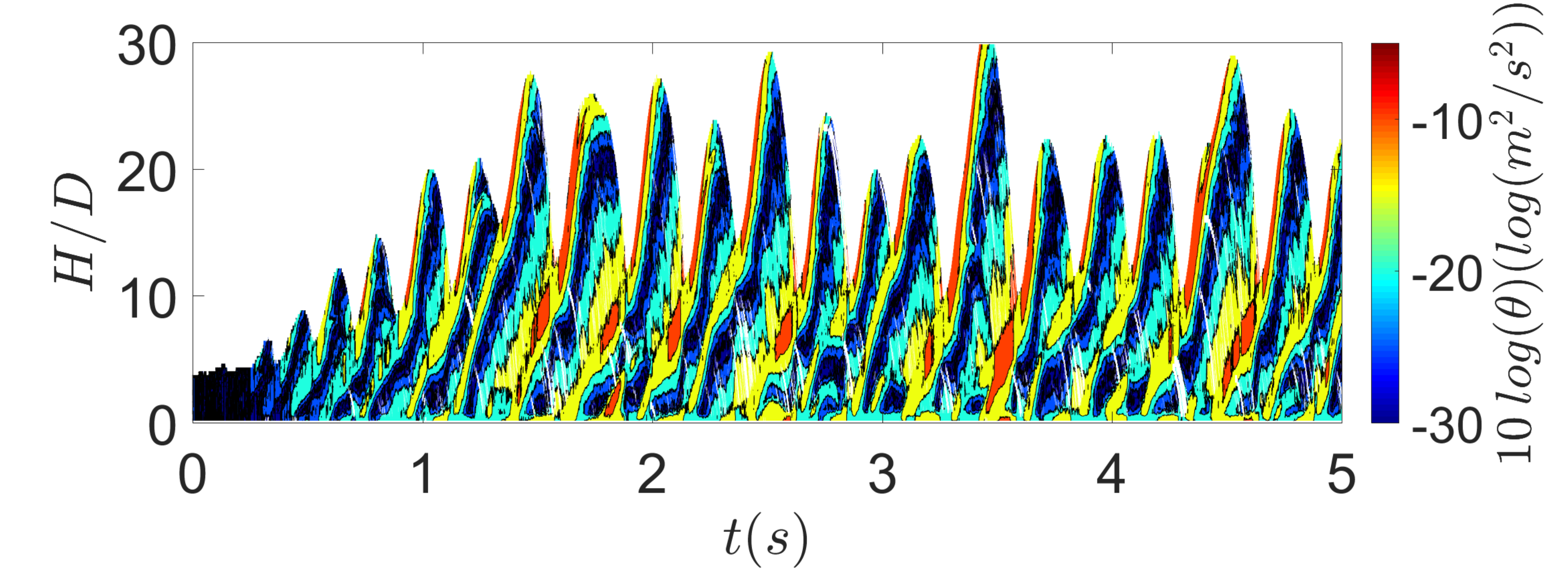}\\
			(b)\\
		\end{tabular} 
	\end{center}
	\caption{Spatio-temporal diagrams of cross-sectional averages of the granular temperature $\theta$ for cases \textit{i} and \textit{k} of Tab. \ref{tab_exp}. The abscissa shows the time $t$, the ordinate the bed height normalized by the tube diameter, $H/D$, and the colors correspond to 10$\log \theta$. Time between frames is of 10$^{-2}$ s and the colorbar is on the right of the figure.}
	\label{fig_gran_temp}
\end{figure}

For each instant, we first measured the transverse $x$ and longitudinal $y$ components of the instantaneous velocity of each grain, $U_p$ and $V_p$, respectively, for each image pair (an example of snapshots showing the instantaneous velocity of each particle is available in the supplementary material). We afterward computed the ensemble average of the velocity considering all particles in the bed, $U_{avg}$ and $V_{avg}$, and next the $x$ and $y$ components of velocity fluctuations, $u_p$ = $U_p - U_{avg}$ and $v_p$ = $V_p - V_{avg}$, respectively, for each image pair. Finally, we computed the instantaneous two-dimensional (2D) granular temperature $\theta_{inst}$ as in Eq. \ref{Eq:grain_temp}, 

\begin{equation}
	\theta_{inst} \,=\, \frac{1}{2} \left( u_p^2+v_p^2 \right) \,\,,
	\label{Eq:grain_temp}
\end{equation}

\noindent where $\theta_{inst}$ is an ensemble-based  granular temperature. In order to analyze the time evolution of the granular agitation along the tube, we computed averages of $\theta_{inst}$ within horizontal slices. For that, we divided the bed in vertical regions (one-dimensional meshes) wherein we computed the ensemble average of the granular temperature. For the 2D data, this is the equivalent of instantaneous values of cross-sectional averages of the granular temperature. Some examples can be seen in Fig. \ref{fig_gran_temp}, which shows spatio-temporal diagrams of cross-sectional averages of the granular temperature, $\theta$, for cases \textit{i} and \textit{k}. In the figure, we plotted 10$\log \theta$ (instead of $\theta$) in order to accentuate differences. We observe first a much lower level of agitation in case \textit{i} than in case \textit{k}, the latter having larger bubbles and smaller plugs. The same increase in the granular temperature with increasing the air flow is observed for the other cases (the graphics are available in the Supplementary Material).

\begin{figure}	
	\begin{center}
		\begin{minipage}{0.59\linewidth}
			\begin{tabular}{c}
				\includegraphics[width=0.95\columnwidth]{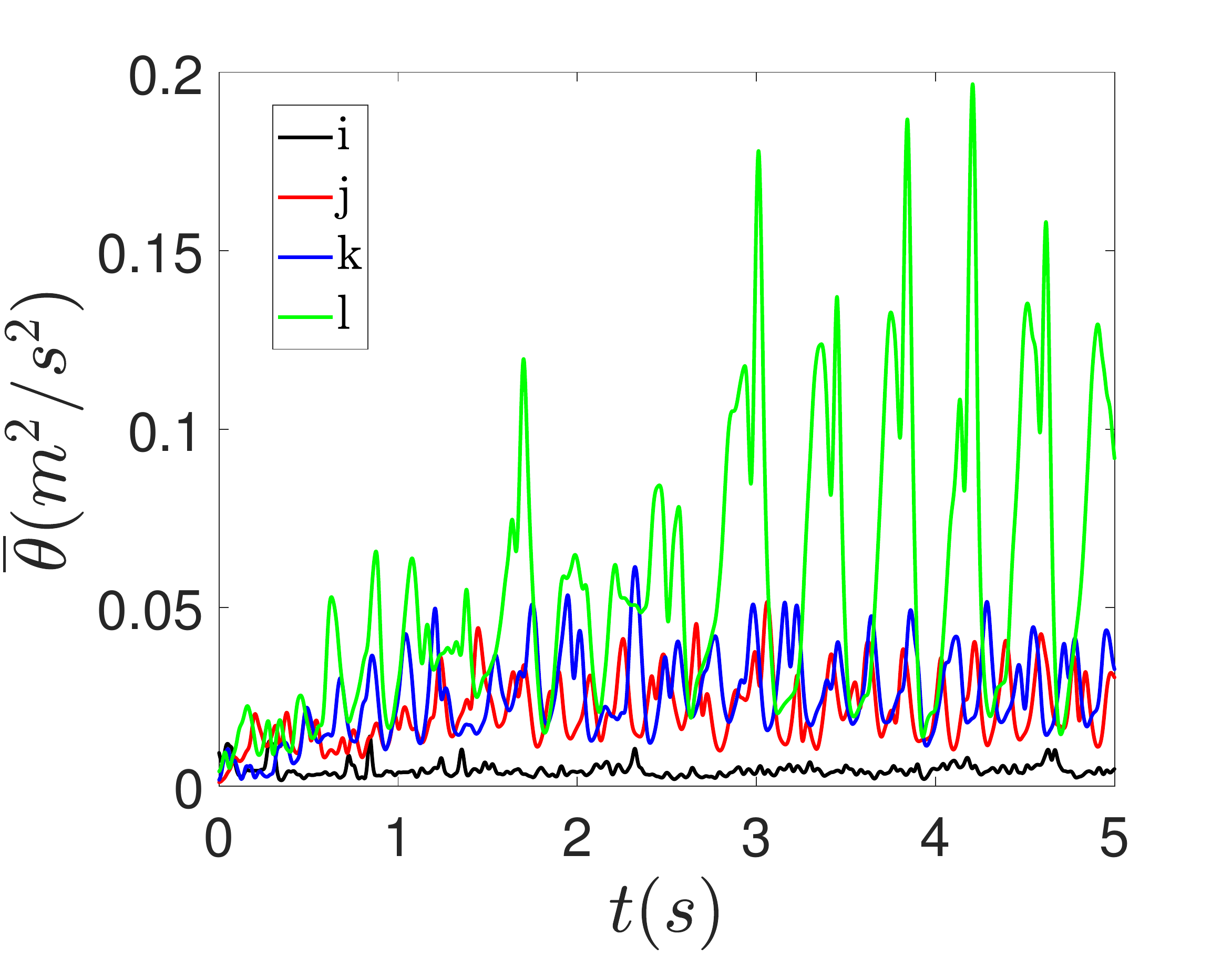}\\
				(a)
			\end{tabular}
		\end{minipage}
		\hfill
		\begin{minipage}{0.39\linewidth}
			\begin{tabular}{c}
				\includegraphics[width=0.9\linewidth]{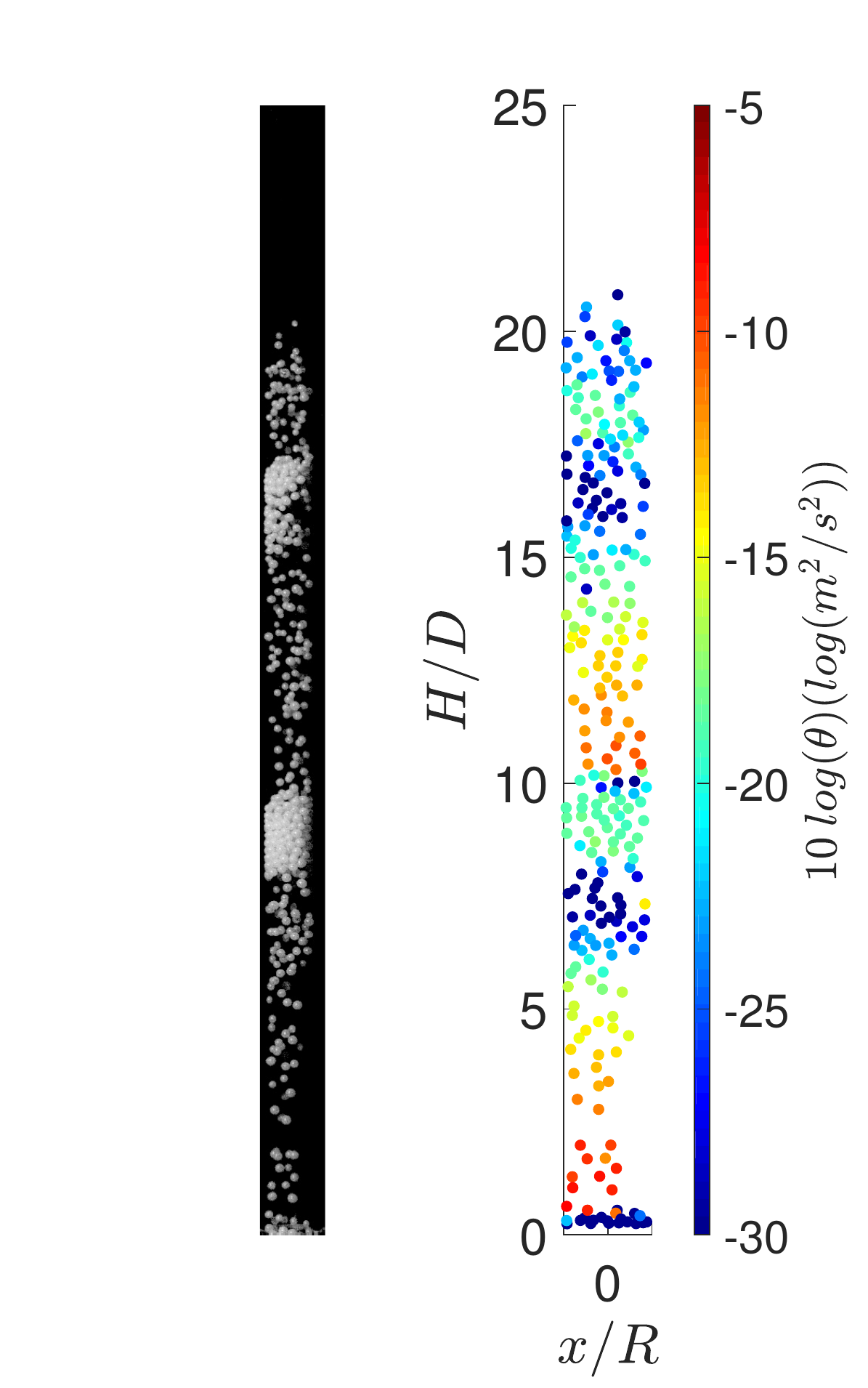}\\
				(b)
			\end{tabular}
		\end{minipage}
		\hfill
	\end{center}
	\caption{(a) Space averaged granular temperature $\overline{\theta}$ as a function of time $t$ for cases \textit{i} to \textit{k} of Tab. \ref{tab_exp}. (b) Example of the granular temperature $\theta_{inst}$ for each grain detected in a given image pair (on the left one of the images of the pair, and on the right $\theta_{inst}$ for each grain).}
	\label{fig_gran_temp_avg}
\end{figure}

We computed also the vertical averages of $\theta$, which correspond to the instantaneous average of the granular temperature for the ensemble of grains, $\overline{\theta}$. Figure \ref{fig_gran_temp_avg}a shows the time evolution of $\overline{\theta}$ for cases \textit{i} to \textit{k} of Table \ref{tab_exp} (a graphic for all cases is available in the Supplementary Material). We notice an increase of $\overline{\theta}$ with $\overline{U}$, with high peaks and oscillations that behave similarly as those of the bed. The oscillations are related to the different levels of granular temperature found in plugs (lower) and bubbles (higher), as shown in Fig. \ref{fig_gran_temp_avg}b for each grain at a given instant. Concerning the phases of $\overline{\theta}$ and $H$, we note that there exists some degree of dispersion given by the motion of parcels of grains and particle-particle shocks (a figure showing both the bed height and the granular temperature as functions of time is available in the supplementary material). In cases where plug-plug collisions occur, we notice also a fluctuation within the large peaks, as can be clearly seen for case \textit{l} in Fig. \ref{fig_gran_temp_avg}. Therefore, the decrease in plug length and the increase in plug celerity and plug-plug collisions promote higher levels of grain agitation.

The increase in $\theta$ by augmenting $\overline{U}$ seems natural, although for very-narrow SLFBs we showed \cite{Cunez3, Cunez4} that this tendency does not necessarily occur within some ranges of $\overline{U}$. Here, besides showing that $\theta$ increases continually with $\overline{U}$ in gas-solid MFBs, we show in detail the mechanics for reaching higher levels of agitation and, thus, higher transfer rates.


\section{Conclusions}
\label{sec:conclusions}

In this paper, we investigated experimentally a mm-scale gas-solid fluidized bed (MFB), which consisted of 0.5-mm-diameter glass particles suspended by an air flow in a 3-mm-ID glass tube. We varied the air velocity and the bed height, and filmed the bed with a high-speed camera. Afterward, we processed the images with a numerical code for tracking both the entire bed and individual particles. Our results are new and show that: (i) instabilities in the form of alternating high- and low-compactness regions (plugs and bubbles) appear in the bed; (ii) by increasing the fluid velocity, the mean height increases (higher expansions); (iii) for higher flow velocities, plugs are smaller and a considerable number of plug-plug collisions occur. With that, two main frequencies of bed oscillation appear at high velocities; (iv) different from very-narrow SLFBs, the MFBs tested in this work did not undergo neither crystallization nor jamming; (v) the granular temperature (and, thus, the microscopic degree of agitation) is much lower within plugs than within bubbles; (vi) the decrease in plug length and the increase in plug celerity and plug-plug collisions promote higher levels of grain agitation; and, finally, (vii) the increase in the flow velocity does not avoid the appearance of plugs, but the granular temperature increases, mitigating the problem. Our results shed light on detailed mechanisms taking place within the miniaturized bed, providing insights for chemical and pharmaceutical processes involving powders.

\section*{Declaration of Competing Interest}

The authors declare no conflict of interest.

\section*{Acknowledgments}

\begin{sloppypar}
The authors are grateful to FAPESP (Grant nos. 2016/18189-0, 2018/14981-7 and 2018/23838-3) and to CNPq (Grant no. 405512/2022-8) for the financial support provided.
\end{sloppypar}




  \bibliographystyle{elsarticle-num} 
  \bibliography{references}






\end{document}